\documentclass{desyprocA4}

\usepackage{svg}
\usepackage[tight]{subfigure}

\usepackage{xspace}
\newcommand{\epem}{\ensuremath{\textrm{e}^+\textrm{e}^-}\xspace}

\graphicspath{{plots/}}

\def\Lumi{\ensuremath{\mathcal{L}}}

\def\Pem{\ensuremath{P_{e^-}}}
\def\Pep{\ensuremath{P_{e^+}}}

\def\sigmaLR{\ensuremath{\sigma_{\mathrm{LR}}}}
\def\sigmaRL{\ensuremath{\sigma_{\mathrm{RL}}}}

\def\fLR{\ensuremath{f_{\mathrm{LR}}}}
\def\fRL{\ensuremath{f_{\mathrm{RL}}}}

\begin{document}


\title{Interplay of Beam Polarisation and Systematic Uncertainties in Electroweak Precision Measurements at Future $e^+ e^-$ Colliders}

\author{Jakob Beyer$^{1,2}$ and Jenny List$^{1}$ \\
on behalf of the Physics and Detector Group of the ILC's International Development Team\\[1ex]
$^{1}$Deutsches Elektronen-Synchrotron DESY, Notkestr. 85, 22607 Hamburg, Germany \\
$^{2}$ Universit\"at Hamburg, Hamburg, Germany}

\desyproc{DESY-22-026, Contribution to the 24th International Spin Symposium}

\doi

\acronym{Contribution to the 24th International Spin Symposium}

\date{\today}

\maketitle

\begin{abstract}Future high-energy \epem colliders will provide some of the most precise tests of the Standard Model. Statistical uncertainties on electroweak precision observables and triple gauge couplings are expected to improve by orders of magnitude over current measurements. This provides a new challenge in accurately assessing and minimizing the impact of experimental systematic uncertainties. Beam polarization may hold a unique potential to isolate and determine the size of systematic effects. So far, studies have mainly focused on the statistical improvements from beam polarisation. This study aims to assess, for the first time, its impact on systematic uncertainties. A combined fit of precision observables, such as chiral fermion couplings and anomalous triple gauge couplings, together with experimental systematic effects is performed on generator-level differential distribution of 2-fermion and 4-fermion final-states. Different configurations of available beam polarisations and luminosities are tested with and without systematic effects, and will be discussed in the context of the existing projections on fermion and gauge boson couplings from detailed experimental simulations.
\end{abstract}

\section{Introduction}
Longitudinally polarised beams are a special feature of Linear \epem Colliders: Both CLIC and ILC foresee a polarisation of the electron beam of about 80\%, as already achieved at the SLC. ILC furthermore aims to polarise also the positron beam to initially 30\%, with an upgrade option to 60\%. Since the sign of the polarisation can be chosen independently for each of the two beams, and since left-handed and right-handed fermions are different particle species under the weak interaction, the ILC can be regarded as being ``four colliders in one'':  For processes mediated by $s$-channel exchange of a vector boson, only the cross-sections for configurations with opposite beam chiralities are non-zero, while $\sigma_{LL}=\sigma_{RR}=0$ by angular momentum conservation. The exchange of a $W$-boson or a neutrino in the $t$-channel is only allowed for left-handed electrons and right-handed positrons, while $\sigma_{RL}=\sigma_{LL}=\sigma_{RR}=0$. Only in the case of $t$-channel photon, $Z$-boson or electron exchange, all four chiral cross-sections are non-zero. The strong dependency of electroweak processes on the beam polarisation enhances the physics potential of an \epem collider by four basic mechanisms: The first three, namely the enhancement of selected signal processes, the suppression of backgrounds and the availability of additional chiral observables has been discussed widely in the literature, see e.g.~\cite{Fujii:2018mli, Moortgat-Pick:2005jsx} for overview articles. This contribution will focus on the forth, maybe less well-known one, namely the role of polarised beams in the control of experimental systematic uncertainties. These can be associated with the measurement of accelerator parameters like the luminosity or the polarisation itself, or with the alignment and calibration of the main detector itself. Many of these effects are time-dependent, and while their long-term average can typically be calibrated very well, their time-variations introduce a spread around this long-term average which usually can only be treated as uncertainty. Now, if -- and only if -- the polarisation signs can be flipped on a time scale much faster than the unavoidable changes of the detector conditions or even configurations, then these effects are strongly correlated between the datasets recorded for the different polarisation configurations and can be treated correspondingly in combined interpretations.
 
\section{Beam Polarisation and Systematic Uncertainties}
As a simplified toy example, let us consider one $s$-channel mediated signal process, one (dominantly) $t$-channel mediated background process and one systematic effect depending on one observable --- for instance $\epem \to \mu^+\mu^-$ or $\epem \to \mu^+\mu^- h$, with two non-zero cross-sections \sigmaLR$^S$ and $\sigmaRL^S$ as signal, $\epem \to (W^+ W^-) \to \mu^+ \nu_{\mu} \mu^- \bar{\nu}_{\mu}$ as main background process with $\sigmaLR^B$ and $\sigmaRL^B$ and the reconstruction efficiency as a function of the polar angle of the $\mu^-$, $\epsilon (\theta_{\mu^-})$. This efficiency can change over time, e.g.\ due to the accelerator-detector alignment, or due to (temporarily) insensitive channels, just to name a few. At any given point in time, the angle-dependent reconstruction efficiency is the same on each data set.
Thus for each data set, the count rate binned as function of the $\mu^-$'s polar angle can be written as:
\begin{equation}
\frac{d N_{i}}{d \cos\theta_{\mu^-}} = \left( \fLR ( \sigmaLR^S \frac{d \sigmaLR^S(i)}{d \cos\theta_{\mu^-}} + \sigmaLR^B \frac{d \sigmaLR^B(i)}{d \cos\theta_{\mu^-}})
+ \fRL ( \sigmaRL^S \frac{d \sigmaRL^S(i)}{d \cos\theta_{\mu^-}}  + \sigmaRL^B \frac{d \sigmaRL^B(i)}{d \cos\theta_{\mu^-}} ) \right) \cdot \epsilon_{i} 
\end{equation}  
Here the $f's$ encode the polarisation and luminosity dependence as $\fLR = (1-\Pem)(1+\Pep) \times \Lumi$, $\fRL = (1+\Pem)(1-\Pep) \times \Lumi$, and $\frac{d \sigma(i)}{d \cos\theta_{\mu^-}}$ denotes differential cross-section in bin $i$, normalised to unity integral. In our simplified example, we will assume that these differential shapes are -- apart from $\epsilon_{i}$ -- perfectly known. In reality, there would of course be more parameters of interest related to the shape, like forward-backward asymmetries etc.

For the case of no beam polarisation 
there is only one data set, 
with $n$ observables from one count rate per bin, which are not enough to determine the $n$ $\epsilon_{i}$ and the two total unpolarised cross-sections $\sigma_0^S$ and $\sigma_0^B$,
meaning that the $\epsilon_{i}$ need to be replaced by an efficiency model with less free parameters, or they have to be determined from simulations or control samples. If at least one beam is polarised 
we have two 
data sets, which --- for fast helicity reversal --- share the same $\epsilon_{i}$, but have different total cross-sections for signal and background. Thus, $2n$ observables are available to determine $n + 4$ unknowns, which gives a fully constrained system for $n \ge 4$, i.e.\ at least four bins. In order to avoid additional uncertainties from finite knowledge of the beam polarisations, their value can be constrained as well if there are at least eight bins. When both beams are polarised, even four data sets, i.e.\ $4n$ observables, are available to constrain $n+4$ or $n+8$ unknowns, resulting in overconstraining for at least two or three bins, respectively. In either of the polarised cases, external information from simulations, model assumptions and control regions could still be used in addition, but is not obligatory anymore.

It should benoted that a global bias in all $\epsilon_{i}$ can be distinguished from modifications in all polarised cross sections. From a physics point of view, it seems very unlikely that all cross sections of various different physics processes (in reality can use even more than the two of our toy example) are off from their SM predictions in exactly the same way. Technically, theory predictions and their uncertainties can be employed as additional constraints in a global fit to data from different channels. Then, compensating a bias in $\epsilon_{i}$ by modified polarised cross sections will give an extra $\chi^2$ penalty for each included process. Another question is whether also the level of residual instrumental backgrounds can be determined. For the sake of the argument, let's follow again the brute-force approach of one ``fake rate'' per bin, without relying on external knowledge. For ony one beam polarised, i.e.\ two data sets one would end up with $2n$ observables to determine $2n + 7$ unknowns, which is never overconstrained, even before adding physics parameters influencing the shape of the distribution. With both beam polarised, however, $4n$ observables to determine $2n + 8$ unknowns, which is overconstraining for more than 2 bins. Many experimental effects will be time-dependent. Data collected at a macroscopically different time (a month later, a year later, ...) needs to be corrected for these effects --- as far as the change with time can be determined. Any residual deviation from or jittering around the best estimate of time-dependent effects will lead to a finite minimum of systematic uncertainties. Concurrently taken data sets with different beam polarisation will push this minimum further down.

\section{Towards a Combined Analysis of $e^+e^-\to f \bar{f}$ and $e^+e^-\to 4f$ Including Systematic Uncertainties}
As a first step towards a more realistic application, a combined fit of differential distributions of $e^+e^-\to \mu^+ \mu^-$ and $e^+e^-\to \mu^{\pm} \nu_{\mu} jj$  at $\sqrt{s}=250$\,GeV, based on events generated  with Whizard2~\cite{Kilian:2007gr}, been set up~\cite{Beyer:2021syl, Beyer:2021nbp}. The beam polarisations and the luminosity are treated as nuisance parameters, optionally including constraints from the beam polarimeters and luminosity measurement. For $e^+e^-\to \mu^+ \mu^-$, the cosine of the polar angle of the $\mu^-$ in the restframe of the di-muon system, $\cos{\theta^*}$, is used as observable. The sample is split into a ``high-E'' and a ``return-to-Z'' part. These distributions are described in case of polarised beams by the total unpolarised cross-section $\sigma_0$, the inital- and final-state asymmetries $A_e$ and $A_{\mu}$  and parameters modeling shape effects due to $Z$-$\gamma$-interference, $\epsilon_{\mu}$ and initial-state radiation, $k_0$ and $\delta k$. Of the latter two, $k_0$ is chirality-independent and $\delta k$ chirality-dependent. More importantly, the two asymmetries and the $Z$-$\gamma$-interference cannot be distinguished anymore because they are all linear in $\cos{\theta^*}$, and collapse into a single parameter, $A_{FB,0}$. 
The considered four-fermion final-states originate from $W$ pair production, and consider three observables, namely the cosine of the $W$ production angle $\cos{\theta_W}$, as well as the decay angles of the leptonically decaying $W$ in its restframe, $\cos{\theta_l^*}$ and $\phi_l^*$. The physics parameters are the unpolarised total cross-section $\sigma_0(W)$, the initial-state asymmetry $A_{LR}(W)$ as well as the three anomalous triple gauge couplings in the LEP parametrisation, $g_1^Z$, $\kappa_{\gamma}$ and $\lambda_{\gamma}$. 
As a first detector-related systematic effect, the acceptance of the muon reconstruction was chosen, since it introduced one of the largest systematic uncertainties in the analyses of $e^+e^-\to \mu^+ \mu^-$ at LEP. Instead of the ``brute force'' approach with one acceptance parameter per bin described in the previous section, it was assumed, though, that for detectors at future Higgs factories, the acceptance can be parametrised as a simple ``box'' with a center position (ideally at $\cos{\theta}=0$) and a width. This is motivated by corresponding acceptance studies e.g.\ for the ILC detector concepts~\cite{Behnke:2013lya}. Note that such a two-parameter description of the acceptance is a strong and very likely oversimplifying assumption.

\begin{figure}[htb]
	\centering
		\label{fig:nuis}
		\centering
		\includegraphics[width=0.99\textwidth]{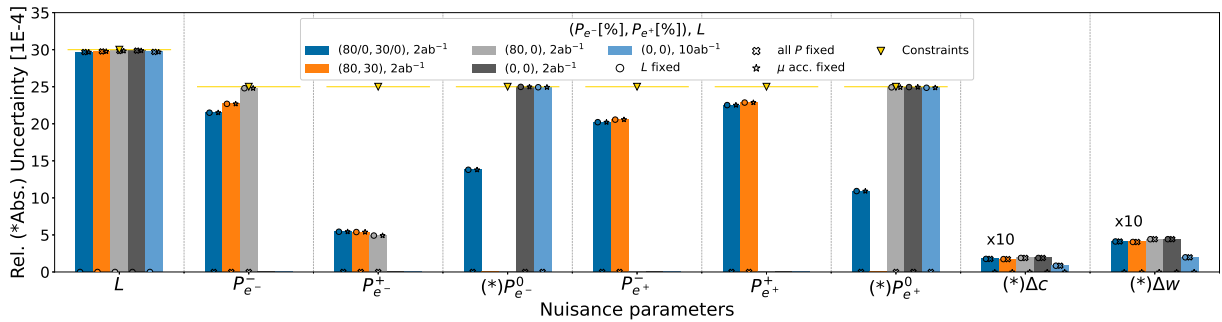}
  
\caption{Precision to which nuisance parameters are determined from the combined $e^+e^-\to \mu^+ \mu^-$ and $e^+e^-\to \mu^{\pm} \nu_{\mu} jj$ fit under different assumptions on the integrated luminosity and the beam polarisations. The yellow triangles with horizontal lines indicate the level of external constraints from the luminosity measurement and the polarimeters.}
\end{figure}

The achievable precisions on the nuisance parameters, i.e.\ the luminosity, the polarisations and the acceptance parameters are displayed in Fig.~\ref{fig:nuis} for a variety of different assumptions on the integrated luminosity and the beam polarisations. The two acceptance parameters are determined at the level of a few $10^{-5}$ (absolute) in this simplistic approach, dominated by the information from $e^+e^-\to \mu^+ \mu^-$. Since the total cross-sections are left as free parameters, the luminosity is purely determined by the assumed precision of the luminosity measurement from low-angle Bhabha scattering. The beam polarisations are determined to precisions between $0.05\%$ and $0.25\%$, almost exclusively relying on information from $e^+e^-\to \mu^{\pm} \nu_{\mu} jj$. If both beams are polarised for at least part of the data, all four (or even six) polarisation values can be determined independently of each other and to better precision than given by the polarimeter constraints. If only one beam can be polarised, e.g.\ the $e^-$ beam, only one out of the three independent polarisation values, namely for the case of a positively polarised $e^-$ beam, can be determined beyond the polarimeter constraint, while the negative $e^-$ polarisation and the $e^+$ polarisation cannot be constrained from the collision data. In case of both beams being unpolarised, neither of the close-to-zero polarisation values can be constrained from collision data, but relies exclusively on polarimeter information or a zero assumption. 

\begin{figure}[tbh]
	\centering
	\hspace{-10pt}
	\subfigure[][]{
		\label{fig:phys:a}
		\centering
\includegraphics[width=0.61\textwidth]{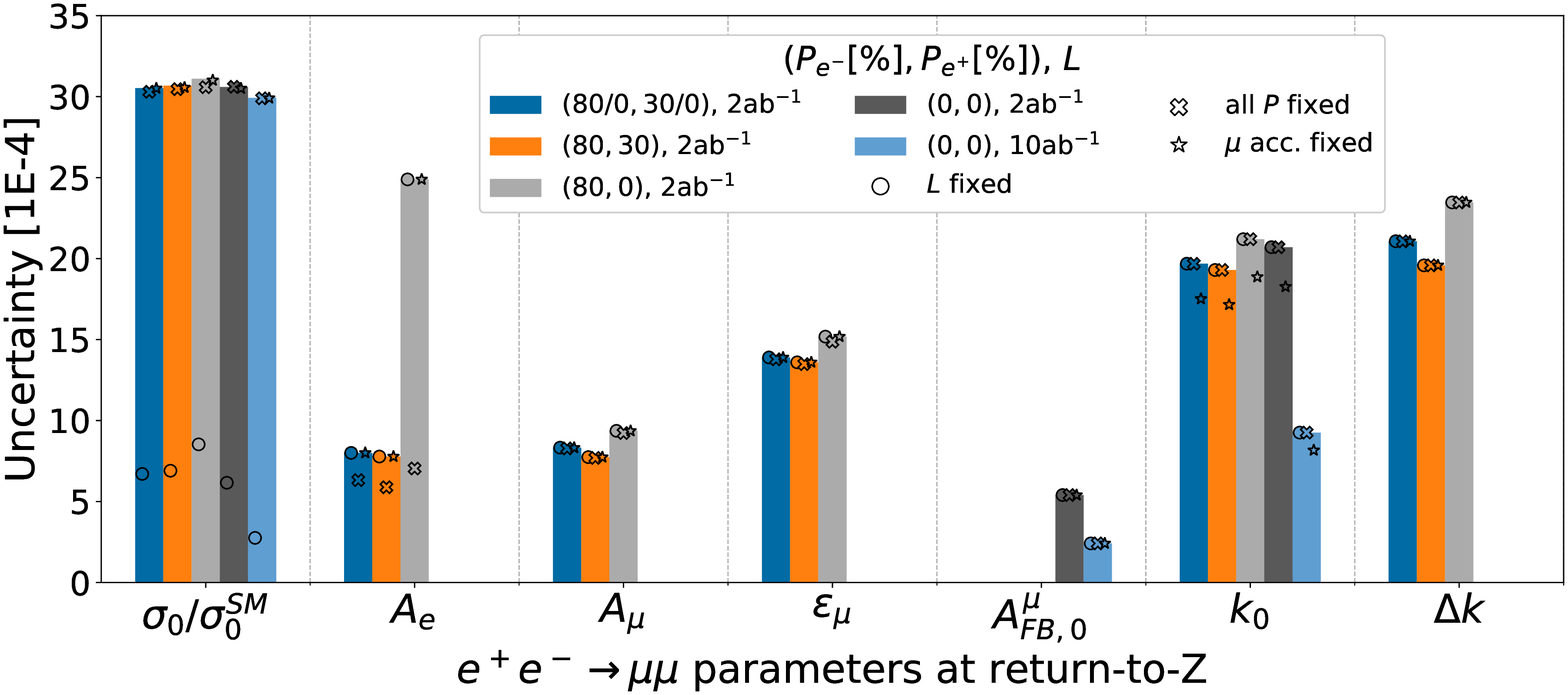}
	}	
	\hspace{-10pt}
	\subfigure[][]{
		\label{fig:phys:b}
		\centering
  \includegraphics[width=0.37\textwidth]{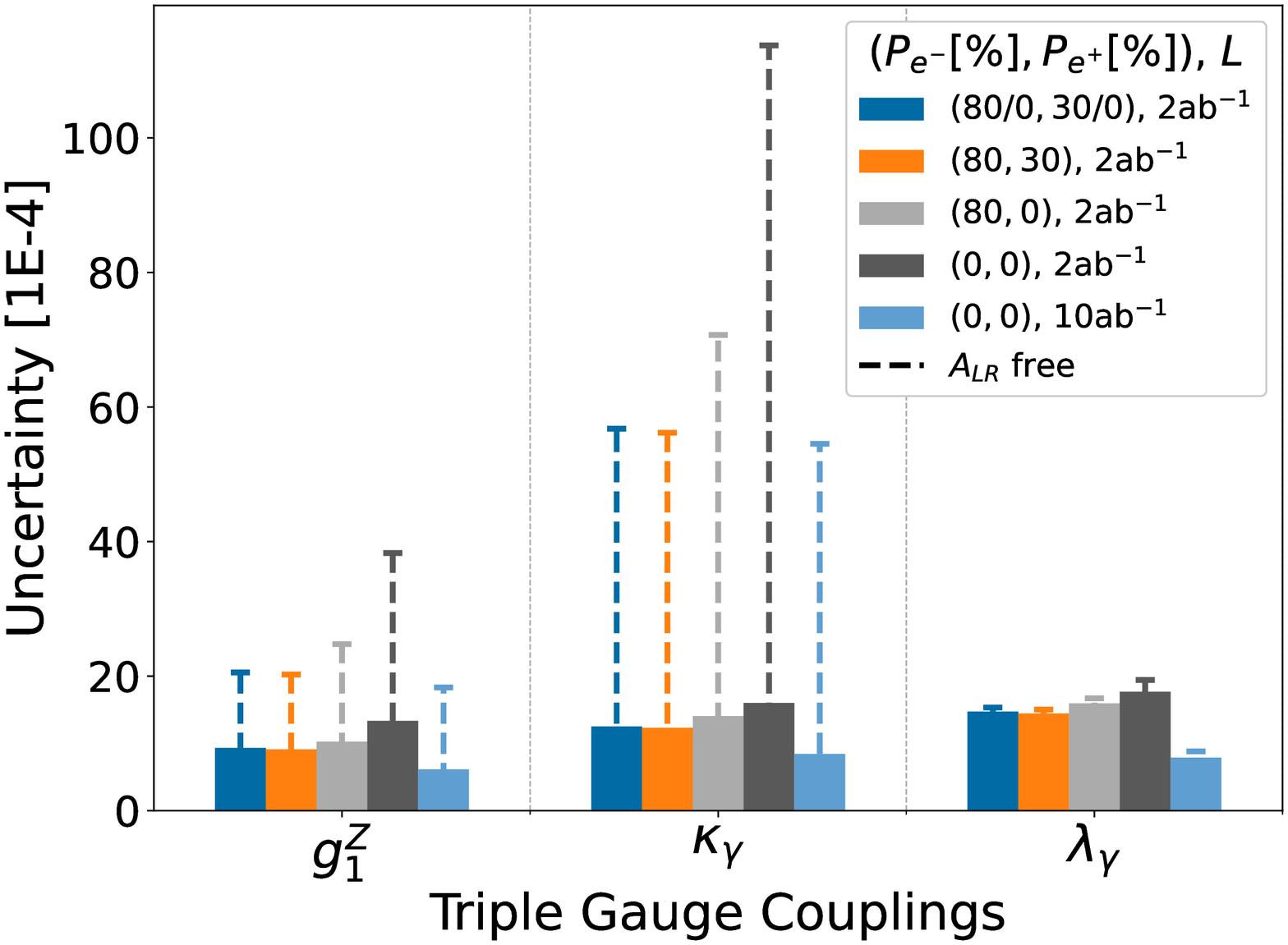}
	}\\
	
\caption{Precisions of physics parameters under different assumptions on the integrated luminosity and the beam polarisations. (a) di-fermion asymmetries and related parameters (b) anomalous triple gauge couplings.}
\end{figure}

The precisions to which the physics parameters of the return-to-Z data set can be determined from the muon channel are shown in Fig.~\ref{fig:phys:a}. If both beams are polarised, the electron asymmetry $A_e$ can be determined to the level of $7 \times 10^{-4}$ with $2$\,ab$^{-1}$, using the muon channel only. This uncertainty receives a visible but small contribution from the polarisation uncertainty, but shows no residual effect of the muon acceptance (various open symbols). If only the $e^-$ beams is polarised, the uncertainty increases by more than a factor of three. The open symbols indicate that this is due to the much worse knowledge of the beam polarisations in this case, while there is still no degradation due to the the muon acceptance. The degradation in $A_{\mu}$ is much smaller. If both beams are unpolarised, $A_e$, $A_{\mu}$ and $\epsilon_{\mu}$ cannot be determined individually, but combine into $A_{FB,0}$ as single parameter, which can be determined to the level of $6 \times 10^{-4}$ only with $10$\,ab$^{-1}$. 

Projections for constraining anomalous triple gauge couplings at future $e^+e^-$ colliders have been derived previously in full detector simulation at $\sqrt{s}=500$\,GeV~\cite{Marchesini:2011aka}, and have been extrapolated by cross-section scaling and a conservative estimation of changes of the detector acceptance to $\sqrt{s}=250$\,GeV~\cite{Karl:2019hes}. For $2$\,ab$^{-1}$ and with both beams polarised, precisions of about $8...10 \times 10^{-4}$ have been obtained from this extrapolation, based on the $\mu^{\pm}\nu_{\mu} jj$ and $e^{\pm}\nu_{e} jj$ channels. This compares well to the result of our fit shown in Fig.~\ref{fig:phys:b}, given that this study does not (yet) include the electron channel, but also neglects selection efficiencies beyond the muon acceptance. In the default setup, the left-right asymmetry of $W$ pair production is fixed to its SM value, since it is strongly correlated with $A_e$ (at the appropriate energy scale). If this constraint is relaxed, and $A_{LR}(W)$ is treated as a free parameters, the uncertainties on $g_1^Z$ and $\kappa_{\gamma}$ increase substantially, as indicated by the dashed bars in Fig.~\ref{fig:phys:b}. This deterioration of the precision, which has also been observed in hadron collider projections~\cite{Azzi:2019yne}, is even more striking in the case of unpolarised beams, and underlines the need to know $A_e$ precisely, also at higher energy scales, and to persue combined interpretations of various physics processes, be it EFT-based approaches or specific UV-complete models.
In all cases, the residual effect of  muon acceptance seems negligible, at least in the very simple ``box-like'' acceptance model assumed so far. More complex and realistic acceptance models should be tested before drawing final conclusions.

\section{Conclusions}
While the role of beam polarisation for chiral analysis, signal-to-background ratio and effective luminosity at future $e^+e^-$ colliders has been studied since many years, its interplay with experimental systematic uncertainties is only starting to be explored. In this study, first steps have been undertaken to consider experimental systematics in a combined fit of $e^+e^-\to f\bar{f}$ and $e^+e^- \to 4 f$ processes. It so far showed e.g.\ that positron polarisation is essential to avoid the measurement of the initial state asymmetry $A_e$ being limited by the polarimeter precision, and that a precise knowledge of $A_e$(250\,GeV) will be crucial for a precise determination of triple gauge couplings. In both cases, four polarisation configurations are better than two or just one. In order to complete the picture, especially w.r.t.\ the role of some of the leading uncertainties in the corresponding analyses at LEP, the study should be extended in the future.

\end{document}